\documentstyle[aps,eqsecnum,epsf]{revtex}

\ifx\epsffile\undefined
\def\insertfig#1{}
\else
\def\insertfig#1{{\baselineskip=4pt
\centerline{\epsfxsize=\hsize\epsffile{#1}}}}\fi
\begin{document}

\def\Tr{{\rm Tr}}
\def\N{N_c}
\def\ls{\Lambda \Sigma^0}
\def\ssqr#1#2{{\vbox{\hrule height #2pt
      \hbox{\vrule width #2pt height#1pt \kern#1pt\vrule width #2pt}
      \hrule height #2pt}\kern- #2pt}}
\def\sqr{\mathchoice\ssqr8{.4}\ssqr8{.4}\ssqr{5}{.3}\ssqr{4}{.3}}

\def\bsqr{\ssqr{10}{.1}}
\def\nbox{\hbox{$\bsqr\bsqr\bsqr\bsqr\raise2.7pt\hbox{$\,\cdot\cdot
\cdot\cdot\cdot\,$}\bsqr\bsqr\bsqr$}}
\def\onebox{{\vbox{\hbox{$\sqr\thinspace$}}}}
\def\twobox{{\vbox{\hbox{$\sqr\sqr\thinspace$}}}}
\def\threebox{{\vbox{\hbox{$\sqr\sqr\sqr\thinspace$}}}}

\twocolumn[       
{\tighten
\preprint{\vbox{
\hbox{UCSD/PTH 95--01}
\hbox{hep-ph/95mmnnn}
}}
\title{Baryon Mass Splittings in the 1/$\bf \N$ Expansion}
\author{Elizabeth Jenkins\footnotemark and Richard F. Lebed}
\address{Department of Physics, University of California at
San Diego, La Jolla, CA 92093}
\bigskip
\date{February 1995}
\maketitle
\widetext
\vskip-1.5in
\rightline{\vbox{
\hbox{UCSD/PTH 95--01}
\hbox{hep-ph/9502227}
}}
\vskip1.5in
\begin{abstract}
The $I=0,1,2,3$ mass splittings of the spin-$1/2$ octet and spin-$3/2$
decuplet baryons are analyzed in the $1/\N$ expansion combined with
perturbative flavor breaking.  We show there is considerable
experimental evidence that the baryon masses satisfy the hierarchy
predicted by this expansion.  Since flavor symmetry-breaking
suppression factors alone are not sufficient to describe the observed
hierarchy, we conclude that there is firm evidence for the $1/\N$
expansion in the baryon masses.  Our analysis differs from
non-relativistic $SU(6)$.
\end{abstract}

}

] 

\narrowtext

\footnotetext{${}^*$National Science Foundation Young Investigator.}

\section{Introduction}
	The $1/\N$ expansion has led to new understanding of the
spin-flavor structure of baryons in QCD\cite
{dm,j,bron,djm,lm,cgo,cgkm,lmw,jm,matt,manohar,djmtwo,tsk,ddjm,matt2}.
In the large $\N$ limit\cite{thooft,witten}, it has been shown that
the baryon sector of QCD possesses a contracted spin-flavor
algebra\cite{dm,gervaissakita}.  Corrections to the large $\N$ limit
can be parametrized by $1/\N$-suppressed operators with definite
spin-flavor transformation properties\cite{dm,j}.  By studying the
spin-flavor structure of these $1/\N$ corrections, it is possible to
obtain new symmetry relations which are satisfied to non-trivial
orders in the $1/\N$ expansion, where the accuracy of these relations
is predicted by the expansion.  In the cases studied thus far, the
$1/\N$ expansion has yielded predictions for static properties of
baryons which agree with the experimental data at the level of
precision predicted by the expansion.

	In this paper we analyze the isospin mass splittings of the
spin-$1/2$ octet and spin-$3/2$ decuplet baryons in the $1/\N$
expansion, with isospin symmetry breaking treated perturbatively.
$SU(3)$ flavor symmetry breaking is treated first perturbatively, and
then nonperturbatively through the use of $SU(2)_I
\times U(1)_Y$ flavor symmetry.  We find that there is evidence for the
pattern of mass splittings predicted by the $1/\N$ expansion and
flavor symmetry breaking.  A number of our predictions are not tested
by the experimental data because they involve baryon mass splittings
which are poorly measured; more accurate measurements of these
splittings would test the validity of the expansion.  The analysis we
perform in this work illustrates that the predictions of the $1/\N$
expansion are different from the standard $SU(6)$ predictions of the
non-relativistic quark model\cite{rss,lebed}.

	The analysis of the isospin mass splittings in the $1/\N$
expansion is organized as follows.  Section~II presents the relevant
operator analysis.  $1/\N$ expansions are constructed for the $SU(2)
\times SU(3)$ representations $(0,1)$, $(0,8)$, $(0,27)$, $(0,64)$ and
$(0, 10 + \overline{10})$.  In Sec.~III, we give the complete set of
linearly independent operators which spans the baryon masses for
$\N=3$.  Each operator occurs at a particular order in $1/\N$ and
flavor symmetry breaking.  Mass relations for the octet and decuplet
baryons are derived in Secs.~IV and V, where we separate the relations
into isospin channels $I=0$, $1$, $2$ and $3$.  In Sec.~IV we present
the analysis with perturbative $SU(3)$ flavor symmetry breaking;
Sec.~V repeats the analysis using $SU(2) \times U(1)$ flavor symmetry.
We contrast the results of the two analyses, and comment on their
relation to $SU(6)$ formul\ae.  Our conclusions are presented in
Sec.~VI.
\section{Operator Analysis}
	The lowest-lying baryons for $\N$ colors transform according
to the completely symmetric spin-flavor $SU(2F)$ representation shown
in Fig.~1.  This baryon representation decomposes under (spin
$\otimes$ flavor) into a tower of baryon states with spins $J=1/2,$
$3/2$, $\ldots$, $\N/2$, with the respective flavor representations
displayed in Fig.~2.  In the following analysis, we consider the
special case of $F=3$ light flavors.  For three light flavors, the
flavor representations of Fig.~2 for $\N$ large and finite differ from
the flavor representations for $\N=3$.  The weight diagrams for the
flavor representations of the spin-$1/2$ and spin-$3/2$ baryons are
given in Figs.~3 and~4, respectively.  These flavor representations
reduce to the baryon octet and decuplet for $\N=3$.  For arbitrary
$\N$, the familiar spin-$1/2$ and spin-$3/2$ baryons can be identified
with states at the top of the flavor representations, for which the
number of strange quarks is $O(1)$ (not $O(\N)$).  In the following
analysis, we are only interested in the masses of baryon states which
continue to exist for $\N=3$; we call these states the physical
baryons.  Focusing on these baryons results in a number of
simplifications in the analysis.  We caution the reader that the
$1/\N$ analysis we perform here for arbitrary $\N$ is valid only for
the physical baryons in Figs.~3 and~4, not the entire flavor
representations.  The results are also valid if the number of colors
is set equal to three.

	We construct an operator expansion for the mass splittings of
the baryon octet and decuplet using quark operators as the operator
basis.  Equivalent results can be derived in the Skyrme operator
basis.  The complete classification of quark operators for three
flavors was performed in Ref.~\cite{djmtwo}; in this work we use the
same notation.  The sole zero-body operator is denoted by $\openone$,
and the complete set of irreducible one-body operators is denoted by
\begin{eqnarray}\label{jtg}
J^i & = & q^\dagger \left(J^i \otimes \openone \right) q\qquad (1,1),
\nonumber \\
T^a & = & q^\dagger \left(\openone \otimes T^a \right) q\qquad
(0,8),\label{IIii} \\
G^{ia} & = & q^\dagger \left(J^i \otimes T^a \right) q \qquad (1, 8),
\nonumber
\end{eqnarray}
where $J^i$ are the spin generators, $T^a$ are the flavor generators,
and $G^{ia}$ are the spin-flavor generators.  These operators
transform as irreducible representations of $SU(2) \times SU(3)$,
which are denoted in Eq.~(\ref{jtg}) by the spin $J$ and the dimension
of the $SU(3)$ flavor representation.

	The $1/\N$ expansion of any operator transforming according to
a given $SU(2) \times SU(3)$ representation is given by an expansion
of the form
\begin{equation}\label{genexp}
{\cal O} = \sum_n c_n {1 \over \N^{n-1}}{\cal O}_n,
\end{equation}
where the $n$-body operators ${\cal O}_n$ are of the generic form
\begin{equation}
{1 \over \N^{n-1}}{\cal O}_n = N_c \left( \frac{J^i}{\N} \right)^l
\left( \frac{T^a}{\N} \right)^m \left( \frac{G^{jb}}{\N}
\right)^{n-l-m} ,
\end{equation}
and the $c_n$ are unknown coefficients.  The number of operators
participating in the expansion~(\ref{genexp}) can be reduced to a
finite set using operator identities.  The operator classification of
Ref.~\cite{djmtwo} showed that the complete set of linearly
independent quark operators for the baryon representation Fig.~1
transforms according to the irreducible spin-flavor representations
\begin{equation}\label{sftensors}
1 + T^\alpha_\beta + T^{(\alpha_1\alpha_2)}_{(\beta_1\beta_2)} +\ldots
+ T^{(\alpha_1\alpha_2\ldots\alpha_{N_c})}_
{(\beta_1\beta_2\ldots\beta_{N_c})},
\end{equation}
where
$T^{(\alpha_1\alpha_2\ldots\alpha_m)}_{(\beta_1\beta_2\ldots\beta_m)}$
is a traceless tensor completely symmetric in its $m$ upper and lower
indices.  Ref.~\cite{djmtwo} proved that the spin-flavor
representation
$T^{(\alpha_1\alpha_2\ldots\alpha_m)}_{(\beta_1\beta_2\ldots\beta_m)}$
consists of purely $m$-body quark operators, since all $n$-body quark
operators $(n>m)$ transforming according to this tensor representation
can be reduced to $m$-body operators using non-trivial operator
identities.  The expansion~(\ref{sftensors}) terminates at $\N$-body
operators, since no higher than $\N$-body operators are required to
describe any spin-flavor operator acting on an $\N$-quark baryon.

	The above analysis implies that the complete $1/\N$ expansion
of any operator transforming according to a given $SU(2) \times SU(3)$
representation can be written in terms of pure $n$-body operators,
which transform according to definite $SU(6)$ representations.  This
set of operators is not the natural basis which arises in an expansion
in flavor symmetry breaking, however.  Instead, the natural basis
consists of operators which are associated with definite powers of
flavor symmetry-breaking parameters; such operators have no contracted
flavor indices.  The operators in this new basis are not pure $n$-body
operators, but contain components which can be reduced to lower-body
operators using non-trivial operator identities.  Thus, the operators
which are natural with regard to the flavor-breaking analysis do not
always correspond to definite $SU(6)$ representations, and the results
we obtain differ from $SU(6)$ formul\ae in some instances.

	In the following analysis, we are only interested in the
predictions of the $1/\N$ expansion for the baryons that exist for
$\N=3$, namely the spin-$1/2$ octet and spin-$3/2$ decuplet baryons.
When the set of baryon states considered for large $\N$ is restricted
to these physical baryons, all $n$-body operators with $n>3$ are
redundant and linearly dependent on $0$-, $1$-, $2$- and $3$-body
operators.  Thus, the complete set of independent operators acting on
this restricted set of baryon states for any $\N$ is given by the
spin-flavor representations
\begin{eqnarray}\label{sfthree}
&&\left( \overline{\threebox} \otimes \threebox \right)
\nonumber \\
&&=  1 + T^\alpha_\beta + T^{(\alpha_1 \alpha_2)}_{(\beta_1 \beta_2)}
 + T^{(\alpha_1\alpha_2\alpha_3)}_{(\beta_1\beta_2\beta_3)}
\nonumber \\
&&= 1 + 35 + 405 + 2695,
\end{eqnarray}
where the dimensions of the $SU(6)$ representations are given in the
last line of Eq.~(\ref{sfthree}).

	To analyze the mass splittings of the octet and decuplet
baryons, we need the spin-zero $SU(2) \times SU(3)$ representations of
the quark operators contained in the $SU(6)$ representations $1$,
$35$, $405$, and $2695$.  As is well known, the $1$ contains a
$(0,1)$; the $35$ contains a $(0,8)$; the $405$ contains $(0,1)$,
$(0,8)$ and $(0,27)$; and the $2695$ contains $(0,8)$, $(0,27)$,
$(0,64)$ and $(0, 10 + \overline{10})$\footnote{The representation
$(0, 10 -\overline{10})$ is not allowed by time reversal invariance.}.
If we were interested in $n$-body operators for $n>3$, there would be
additional $SU(3)$ representations to consider.  For example, the
purely $4$-body $SU(6)$ representation contains $(0,1)$, $(0,8)$,
$(0,27)$, $(0,27)$, $(0,64)$, $(0,125)$ and $(0, 35 + \overline{35})$.
All of these quark operators either vanish or reduce to $0$-, $1$-,
$2$- and $3$-body operators {\it when one restricts the set of baryons
of interest to the physical baryon states}.

	Thus, in the analysis of the isospin splittings of the octet
and decuplet baryons, we need to construct $1/\N$ expansions only for
the $SU(2) \times SU(3)$ representations $(0,1)$, $(0,8)$, $(0,27)$,
$(0,64)$, and $(0, 10 + \overline{10})$.  Furthermore, the $1/\N$
expansions can be truncated at three-body operators when one considers
only physical baryon states.  The set of $0$-, $1$-, $2$- and $3$-body
operators in the $1/\N$ expansions spans all of the mass splittings of
the physical baryons.  Only if one is able to truncate the $1/\N$
expansions before the occurrence of three-body operators does one
obtain any non-trivial mass relations which are valid to a given order
in the $1/\N$ expansion.  $1/\N$ operator expansions up to $3$-body
operators are obtained for the $(0,1)$, $(0,8)$, $(0,27)$, $(0,64)$,
and $(0, 10 + \overline{10})$ representations in the following
subsections.
\subsection{ $(0,1)$ }
	The complete set of $(0,1)$ operators can be classified using
the operator identities of Ref.~\cite{djmtwo}.  There is only one
zero-body operator transforming as $(0,1)$ under $SU(2) \times SU(3)$,
\begin{equation}
{\cal O}_0 = \openone,
\end{equation}
and only one two-body operator,
\begin{equation}
{\cal O}_2 = J^2 \ .
\end{equation}
In general, there is a single $n$-body operator for each even $n$,
\begin{equation}
{\cal O}_n = J^n \ ,
\end{equation}
which transforms as $(0,1)$.  Thus, the general $1/\N$ expansion for a
$(0,1)$ operator is of the form\cite{j,djm,lm,cgo,djmtwo}.
\begin{equation}\label{1exp}
{\cal O} = \sum_{n=0,2}^{\N-1} c_n {1 \over \N^{n-1}} {\cal O}_n \ ,
\end{equation}
for $\N$ odd, where $c_n$ are unknown coefficients.  Since the matrix
elements of $J$ are order one for baryons with spins at the bottom of
the tower~\cite{dm}, each succeeding operator is suppressed by
relative order $1/\N^2$, and so it is possible to truncate the $1/\N$
expansion~(\ref{1exp}) at any desired point in the expansion.
\subsection{ $(0,8)$ }
	The $1/\N$ expansion for a $(0, 8)$ operator appears in
Ref.~\cite{djmtwo}.  Operator reduction identities imply that only
$n$-body operators with a single factor of either $T^a$ or $G^{ia}$
need to be retained.  There is only one one-body operator,
\begin{equation}\label{adjonebody}
{\cal O}^a_1 = T^a,
\end{equation}
and only one two-body operator,
\begin{equation}\label{adjtwobody}
{\cal O}^a_2 = \left\{ J^i, G^{ia} \right\},
\end{equation}
allowed after operator reduction.  In general, there is only one
independent $n$-body operator for each $n$.  All of these operators
can be generated recursively from ${\cal O}^a_1$ and~${\cal O}^a_2$ by
anticommuting with $J^2$,
\begin{equation}
{\cal O}^a_{n+2} = \left\{J^2, {\cal O}^a_n \right\}\ .
\end{equation}
The set of operators ${\cal O}^a_n$, $n=1,2,\ldots,\N$, forms a
complete set of linearly independent spin-zero octets.  Thus, the
$1/\N$ expansion for a $(0,8)$ has the form
\begin{equation}\label{8exp}
{\cal O}^a=\sum_{n=1,2}^{\N} c_n {1\over \N^{n-1}} {\cal O}^a_n,
\end{equation}
where $c_n$ are unknown coefficients.  Since $J$ is of order one, the
operator ${\cal O}^a_{n+2}$ in Eq.~(\ref{8exp}) is suppressed by
$1/\N^2$ relative to ${\cal O}^a_n$.  Thus, it is valid to truncate
the expansion~(\ref{8exp}) for arbitrary $a$ after the first two
terms, up to corrections of relative order $1/\N^2$.  Since for
$\N=3$, the expansion Eq.~(\ref{8exp}) ends with the three-body
operator
\begin{equation}
{\cal O}^a_3 = \{ J^2, T^a \},
\end{equation}
this truncation amounts to the neglect of one operator.

	There are two different $(0,8)$ operators which are relevant
for the analysis of the baryon mass splittings: ${\cal O}^8$ with
$I=0$ and ${\cal O}^3$ with $I=1$.  It is important to further analyze
the implicit $\N$-dependence of the matrix elements of ${\cal O}^a_1$
and ${\cal O}^a_2$ for the physical baryons in these special cases.
The matrix elements of the one-body operators $T^a$ and $G^{ia}$,
$a=8$ and $a=3$, can be rewritten in terms of quark number and spin
operators,
\begin{eqnarray}\label{tgid}
&&T^8 = \frac{1}{2\sqrt{3}} \left( N_c - 3N_s \right) ,\nonumber\\
&&G^{i8} = \frac 1 {2\sqrt{3}} \left( J^i - 3 J_s^i \right),
\nonumber\\
&&T^3 = \frac 1 2 \left(N_u - N_d \right),\\
&&G^{i3} = \frac 1 2 \left( J_u^i - J_d^i \right),\nonumber
\end{eqnarray}
where $\N= N_u + N_d +N_s$ and $J^i = J_u^i + J_d^i + J_s^i$.  Using
Eq.~(\ref{tgid}), one finds the matrix elements
\begin{eqnarray}
\{ J^i , G^{i8} \}  &&= \frac{1}{2\sqrt{3}} \left( 2 J^2 - 3
\{J^i , J_s^i \} \right) \nonumber\\
&&= \frac{1}{2\sqrt{3}} \left( - J^2 + 3 I^2 -3 J_s^2 \right),
\nonumber\\
\{ J^i , G^{i3} \}  &&= \frac{1}{2} \left( \{J^i , J_u^i \} -
\{J^i , J_d^i \}   \right) \\
&&= \frac{1}{2} \left( V^2 - U^2 + J_u^2 - J_d^2 \right),\nonumber
\end{eqnarray}
where $I$, $U$ and $V$ are the isotopic spins of $SU(3)$.  The baryon
states of physical interest are those with spin, isospin and
strangeness of order unity.  Thus, these baryons have matrix elements
of $T^8$, $\{ J^i, G^{i8} \}$, $T^3$ and $\{ J^i, G^{i3}\}$ which are
$O(\N)$, $O(1)$, $O(1)$ and $O(\N)$\footnote{From Figs.~3 and~4, it is
clear that the physical baryons have $J_u$, $J_d$, $U$-spin and
$V$-spin of $O(\N)$.  The $O(N_c^2)$ contribution to $\{J^i, G^{i3}
\}$ cancels exactly, but the $O(\N)$ piece does not.}, respectively.
The $\N$-dependence of $T^8$ is trivial; the coefficient of the
$O(\N)$ piece in $T^8$ matrix elements is the same for all the
physical baryons, so it cancels in any mass difference.  Furthermore,
consider any operator ${\cal O}_m/N_c^{m-1}$.  Then the part of $\{
T^8, {\cal O}_m \}/N_c^{m}$ which originates in the $O(\N)$ piece of
$T^8$ is of the form
\begin{equation}
{\rm (const)} N_c \times {\cal O}_m/N_c^{m} ,
\end{equation}
and can be absorbed into the operator ${\cal O}_m$.  We therefore do
not need to worry about the $O(\N)$ contribution to the matrix
elements of $T^8$, and we conclude that for the special case $a=8$, it
is possible to truncate the expansion~(\ref{8exp}) after the first
term, $T^8$, up to a correction of relative order $1/\N$ arising from
$\{ J^i, G^{i8} \}/\N$.  The situation for $a=3$ does not simplify
since the matrix elements of $T^3$ and $\{ J^i, G^{i3} \}/\N$ are both
order one.  The first truncation of the expansion for $a=3$ retains
these two operators.
\subsection{ $(0,27)$ }
	The $1/\N$ expansion for a $(0, 27)$ operator can be
determined using the operator reduction rule of Ref.~\cite{djmtwo}.
There are three two-body operators which transform as a flavor $27$:
spin-zero ($\{T^a, T^b\}$), spin-one ($\{G^{ia}, T^b\} + \{G^{ib},
T^a\}$) and spin-two ($\{G^{ia}, G^{jb} \}+\{G^{ib}, G^{ja} \}$).
Spin-zero $27$ operators can be obtained from the latter two by
forming tensor products with factors of $J$ to saturate spin indices.
Thus, there is a unique two-body $(0,27)$ operator,
\begin{equation}\label{sstwobody}
{\cal O}^{ab}_2 = \{ T^a, T^b \} \ .
\end{equation}
There is one three-body $(0,27)$ operator, which is the tensor product
of the spin-one $27$ two-body operator and $J^i$,
\begin{eqnarray}\label{ssthreebody}
{\cal O}^{ab}_3 &&= \{ J^i, \{ T^a, G^{ib} \} \}
+ \{ J^i, \{T^b, G^{ia} \} \}\nonumber\\
&&= \{ T^a, \{J^i, G^{ib} \} \}
+ \{ T^b, \{J^i, G^{ia} \} \} ,
\end{eqnarray}
where the second equality follows since $J^i$ and $T^a$ commute.  The
tensor product of the spin-two $27$ and the spin-two combination
$\{J^i, J^j\}$ yields the four-body $(0, 27)$ operator,
\begin{equation}\label{ssfourbody}
{\cal O}^{ab}_4 = \{ \{ J^i, J^j \}, \{ G^{ia}, G^{jb} \} \} \ ,
\end{equation}
where projection of the spin-two pieces of $\{J^i, J^j\}$ and
$\{G^{ia}, G^{jb}\}$ in Eq.~(\ref{ssfourbody}) is implied.  It is also
to be understood that the flavor singlet and octet components of the
above three operators are subtracted off, so that each of the three
operators is truly a $(0, 27)$.  Note in all these cases the symmetry
under exchange of flavor indices, as required for flavor-$27$
operators.

	In general, the complete set of linearly independent $(0, 27)$
operators consists of three operator series, namely the three
operators ${\cal O}^{ab}_2$, ${\cal O}^{ab}_3$, ${\cal O}^{ab}_4$ and
anticommutators of these operators with $J^2$.  Note that this implies
that there are two different $n$-body operators for $n$ even, $n \ge
4$, since $\{ J^2, {\cal O}^{ab}_{2}\}$ and ${\cal O}^{ab}_4$ are both
four-body operators.  Thus, the $1/\N$ expansion for a $(0,27)$
operator has the form
\begin{equation}\label{27exp}
{\cal O}^{ab} =\sum_{n=2,3}^{\N} c_n
{1\over \N^{n-1}} {\cal O}^{ab}_n
+ \sum_{n=4,6}^{\N-1} d_n
{1\over \N^{n-1}} {\cal O}^{ab}_n,
\end{equation}
where $c_n$ and $d_n$ are unknown coefficients, and the $d_n$
operators are the series of operators generated from ${\cal
O}^{ab}_4$.  Since the matrix elements of $J$ are order one, the
expansion~(\ref{27exp}) can be truncated for arbitrary $a$ after the
first three operators (${\cal O}_2^{ab}$, ${\cal O}_3^{ab}$, ${\cal
O}_4^{ab}$) up to corrections of relative order $1/\N^2$.  For $\N=3$,
one only retains the two operators ${\cal O}_2^{ab}$ and ${\cal
O}_3^{ab}$.  Truncation after ${\cal O}_3^{ab}$ is valid for any $\N$
up to a relative correction of order $1/\N^2$ only if one restricts
the set of baryons of interest to the physical baryons.

	There are three different $(0,27)$ operators which are
relevant for the analysis of the baryon mass splittings: ${\cal
O}^{88}$ with $I=0$, ${\cal O}^{83}={\cal O}^{38}$ with $I=1$, and
${\cal O}^{33}$ with $I=2$\footnote{Subtraction to obtain operators
with a unique value of isospin is understood.}.  For the special case
$a=b=8$, it is possible to truncate the expansion~(\ref{27exp}) after
the first operator in the expansion, ${\cal O}^{88}_2$, up to a
correction of relative order $1/\N$ arising from ${\cal O}^{88}_3/\N$,
since the matrix elements of $\{J^i, G^{i8} \}$ are $O(1)$.  The
truncations for ${\cal O}^{83}$ and ${\cal O}^{33}$ cannot be
simplified further.
\subsection{ $(0,64)$ }
	The $1/\N$ expansion for a $(0,64)$ operator begins with a
single three-body operator,
\begin{equation}
{\cal O}^{abc}_3 = \{ T^a, \{ T^b , T^c \} \} \ ,
\end{equation}
where it is understood that the singlet, octet and $27$ components of
the above operator are to be subtracted off so that only the flavor
$64$ component remains.  For $\N=3$, this is the single operator which
enters the analysis.  For arbitrary $\N$, there are three additional
operators,
\begin{eqnarray}
&&{\cal O}^{abc}_4 = \{ T^a, \{ T^b , \{J^i, G^{ic} \} \} \},
\nonumber \\
&&{\cal O}^{abc}_5 = \{ T^a, \{ \{J^i, G^{ib} \} ,
\{J^i, G^{ic} \} \} \}, \\
&&{\cal O}^{abc}_6 = \{ \{J^i, G^{ia} \}, \{ \{J^i, G^{ib} \} ,
\{J^i, G^{ic} \} \} \}. \nonumber
\end{eqnarray}
All other operators of the expansion are generated from the above four
by anticommuting with $J^2$.  Truncation of the $1/\N$ expansion after
the first operator ${\cal O}_3^{abc}$ is valid for general $\N$ up to
a relative correction of order $1/\N^2$ only if one restricts the set
of baryons of interest to the physical baryons.

	There are four different $(0,64)$ operators which are relevant
for the analysis of the baryon mass splittings: ${\cal O}^{888}$ with
$I=0$, ${\cal O}^{883}$ with $I=1$, ${\cal O}^{833}$ with $I=2$, and
${\cal O}^{333}$ with $I=3$.
\subsection{ $(0, 10 + \overline{10} )$ }
	The $1/\N$ expansion for a $(0, 10 + \overline{10})$ operator
begins with a single three-body operator,
\begin{equation}
{\cal O}^{ab}_3 = \{ T^a, \{J^i, G^{ib} \} \}
- \{ T^b, \{J^i, G^{ia} \} \} .
\end{equation}
For $\N=3$, this is the only operator which enters the analysis.  For
arbitrary $\N$, there are additional operators generated by
anticommutators of $J^2$ with the above operator.  Thus, the
truncation of the general $1/\N$ expansion of a $(0, 10 +
\overline{10})$ operator after this first operator is valid up to a
relative correction of order $1/\N^2$.

	There is one $(0, 10 + \overline{10})$ operator which is
relevant for the analysis of the baryon mass splittings: ${\cal
O}^{38}$ with $I=1$.
\section{Complete Set of Operators}
	In this section, we present the complete set of operators
which parametrize the baryon masses for $\N=3$.  The set of operators
given here are also the operators required for an analysis for
arbitrary $\N$ when one restricts the baryon states of interest to the
physical baryons.  There are 19 Hermitian operators which span the
space parametrized by the 18 baryon masses and the single off-diagonal
mixing mass $\bar \Lambda \Sigma^0 = \bar \Sigma^0 \Lambda$.  Since
$\bar \Lambda \Sigma^0 = \bar \Sigma^0 \Lambda$ by hermiticity, we
refer to the off-diagonal mass as $\Lambda \Sigma^0$ in this
work\footnote{A $(\bar \Lambda \Sigma^0 - \bar \Sigma^0 \Lambda)$ mass
difference, corresponding to the representation $(0, 10
-\overline{10})$, is disallowed by time reversal invariance.}.  The 19
operators that we require are obtained by truncating the general
$1/\N$ expansions derived in the previous section at three-body
operators.

	The baryon mass operator $M$ can be written in terms of mass
operators $M^R_I$, which belong to $SU(3)$ representations $R$ with
isospin $I$,
\begin{equation}
M = \sum_{R,I} M^R_I .
\end{equation}
For the restricted set of baryon states, the most general mass
operator is given by
\begin{eqnarray}\label{mreps}
M & = & M^1_0+M^8_0+M^{27}_0+M^{64}_0+M^{10+\overline{10}}_1
\nonumber\\
& & \mbox{}+M^{8}_1 +M^{27}_1+M^{64}_1+M^{27}_2+M^{64}_2+M^{64}_3 \ .
\end{eqnarray}
Each of the mass operators $M^R_I$ in Eq.~(\ref{mreps}) may be
expanded in a $1/\N$ expansion of the form
\begin{equation}
M^{R}_I = \sum_{n=0}^3 c^{R,I}_{(n)} {1 \over \N^{n-1} } {\cal
O}^{R,I}_n,
\end{equation}
where the summation is over all $n$-body operators ${\cal O}^{R,I}_n$
with the same transformation properties under $SU(3)$ and isospin as
$M^R_I$.  The unknown operator coefficients are denoted by
$c_{(n)}^{R,I}$.  The explicit expansions for each of the
representations appearing in Eq.~(\ref{mreps}) are obtained from the
results of Sec.~II and are given below.  The operator expansions
divide into the $I=0$ expansions
\begin{eqnarray}\label{izero}
&&M^1_0 =  c_{(0)}^{1,0}\ \N \openone + c_{(2)}^{1,0}\ {1 \over \N }
J^2, \nonumber \\
&&M^8_0 =  c_{(1)}^{8,0}\ T^8 + c_{(2)}^{8,0}\ {1 \over \N }
\{ J^i, G^{i8} \} + c_{(3)}^{8,0}\ {1 \over \N^2 } \{ J^2, T^8 \},
\nonumber \\
&&M^{27}_0 =   c_{(2)}^{27,0}\ {1 \over \N} \{ T^8, T^8 \}
+  c_{(3)}^{27,0}\ {1 \over \N^2 } \{ T^8, \{ J^i, G^{i8} \}\},
\nonumber \\
&&M^{64}_0 =   c_{(3)}^{64,0} {1 \over \N^2 } \{ T^8, \{ T^8, T^8
\}\};
\end{eqnarray}
the $I=1$ expansions
\begin{eqnarray}\label{ione}
&&M^8_1 =  c_{(1)}^{8,1}\ T^3 + c_{(2)}^{8,1}\ {1 \over \N }
\{ J^i, G^{i3} \} + c_{(3)}^{8,1}\ {1 \over \N^2 } \{ J^2, T^3 \},
\nonumber \\
&&M^{27}_1  = c_{(2)}^{27,1}\ {1 \over \N} \{ T^3, T^8 \}\nonumber\\
&& \mbox{\quad\ \ } + c_{(3)}^{27,1}\ {1 \over \N^2 } \left( \{ T^3,
\{ J^i, G^{i8} \}\} +\{ T^8, \{ J^i, G^{i3} \}\} \right), \nonumber \\
&&M^{64}_1 =   c_{(3)}^{64,1}\ {1 \over \N^2 } \{ T^8,
\{ T^8, T^3 \}\}, \\
&&M^{10 + \overline{10}}_1 \! \! = c_{(3)}^{10 + \overline{10},1}\
\! \! {1 \over \N^2 } \! \left(\{ T^3 \! , \{ J^i, G^{i8} \} \} \!
- \! \{ T^8 \, , \{ J^i, G^{i3} \} \} \right) ; \nonumber
\end{eqnarray}
the $I=2$ expansions
\begin{eqnarray}\label{itwo}
&&M^{27}_2 = c_{(2)}^{27,2}\ {1 \over \N} \{ T^3, T^3 \} +
c_{(3)}^{27,2}\ {1 \over \N^2 } \{ T^3, \{ J^i, G^{i3} \}\} ,
\nonumber \\
&&M^{64}_2 =   c_{(3)}^{64,2}\ {1 \over \N^2 } \{ T^8,
\{ T^3, T^3 \}\} ;
\end{eqnarray}
and the $I=3$ expansion
\begin{equation}\label{ithree}
M^{64}_3 =   c_{(3)}^{64,3}\ {1 \over \N^2 } \{ T^3,
\{ T^3, T^3 \}\} .
\end{equation}
Note that, as in the previous section, the 19 operators appearing in
Eqs.~(\ref{izero}--\ref{ithree}) are to be regarded as subtracted
operators, so that each operator transforms according to the $SU(3)$
and isospin representations stated.  Thus, the flavor $27$ operators
require subtraction of singlet and octet components, the $64$
operators require removal of singlet, octet and $27$ components, and
$10 + \overline{10}$ operator requires removal of an octet component.
One further level of subtraction diagonalizes the operators into
channels of unique isospin $I = 0$,1,2,3.

	It is easy to incorporate flavor symmetry breaking into the
$1/\N$ analysis by associating powers of symmetry-breaking parameters
with each of the coefficients appearing in
Eqs.~(\ref{izero}--\ref{ithree}).  There are two sources of flavor
symmetry breaking.  The first source is the quark mass matrix, which
introduces the perturbations
\begin{equation}
\epsilon {\cal H}^8 + \epsilon^\prime {\cal H}^3,
\end{equation}
where $\epsilon$ is an $SU(3)$-violating parameter and
$\epsilon^\prime$ is an isospin-violating parameter.  The magnitude of
these symmetry-breaking parameters is governed by quark mass
differences divided by the chiral symmetry breaking scale, which is of
order one~GeV.  The symmetry-breaking parameter $\epsilon \sim 0.25$
is comparable to a $1/\N$ effect in QCD, so we must carefully keep
track of all powers of $\epsilon$ in the perturbative analysis.  The
isospin-breaking parameter satisfies $\epsilon^\prime \ll \epsilon$.
A typical isospin mass splitting is order several~MeV, whereas the
overall $O(\N)$ mass of the baryons is about one~GeV.  Thus,
$\epsilon^\prime$ is comparable to an effect of order $1/\N^5$ in QCD.
The $1/\N$ expansions we have constructed contain only a few orders of
$1/\N$, so we only need to consider isospin-breaking effects to linear
order in the isospin-breaking parameter $\epsilon^\prime$.  The second
source of flavor symmetry breaking is the quark charge matrix.
Electromagnetic mass splittings are second order in the quark charge
matrix, and are suppressed by $\alpha_{em}/4\pi$.  These splittings
are typically of order a few~MeV in magnitude, which is comparable to
the isospin mass splittings arising from the quark mass matrix but is
negligible compared to $SU(3)$ mass splittings.  We introduce the
symmetry-breaking parameter $\epsilon^{\prime\prime}$ for the
electromagnetic mass splittings, where $\epsilon^{\prime\prime} \sim
\epsilon^\prime$.  The electromagnetic effects can occur in the
$I=0,1,2$ channels; the $I=0$ contribution can be neglected, and both
the $I=1$ and $2$ contributions are suppressed by
$\epsilon^{\prime\prime}$.  The symmetry-breaking parameters
associated with the mass operators at leading order in flavor symmetry
breaking are listed in Table~I\footnote{Note that the true flavor
suppression of $M^{10+\overline{10}}_1$ comes from terms of $O(e^2
m_s)$\cite{lebedluty} and is hence $\epsilon^{\prime\prime} \epsilon$.
In Table~I we list the naive factor $\epsilon^\prime \epsilon$, which
is equivalent because $\epsilon^{\prime\prime} \sim
\epsilon^{\prime}$.}.

	Finally, we describe in more detail the relation between the
$1/\N$ expansion when it is combined with a perturbative analysis in
flavor symmetry breaking and a pure $SU(6)$ analysis.  Not all of the
mass operators of the $1/\N$ expansion
Eqs.~(\ref{izero}--\ref{ithree}) with perturbative flavor breaking
transform according to unique $SU(6)$ representations.  Thus, it is
not possible to identify the $n$-body label $n=0,1,2,3$ of the
coefficients $c_{(n)}^{R,I}$ with the $1$, $35$, $405$ and $2695$
dimensional representations of $SU(6)$, respectively.  This subtlety
occurs because some of the $n$-body operators written in
Eqs.~(\ref{izero}--\ref{ithree}) are not pure $n$-body $SU(6)$
operators; the operators contain components which reduce by the
identities of Ref.~\cite{djmtwo} to lower-body operators in different
$SU(6)$ representations.  An upper-triangular matrix summarizes the
relation between the pure $n$-body $SU(6)$ (rows) and $n$-body flavor
symmetry-breaking (columns) operator bases of the $1/\N$ expansion:
\begin{equation}\label{matrix}
\begin{array}{cc}
&
\begin{array}{cccc}
0 & 1 & 2 & 3
\end{array} \\
\begin{array}{r}
1 \\ 35 \\ 405 \\ 2695
\end{array}
&
\left(
\begin{array}{cccc}
* & * & * & * \\
0 & * & * & * \\
0 & 0 & * & * \\
0 & 0 & 0 & *
\end{array}
\right)
\end{array} ,
\end{equation}
where $*$ indicates an entry which is not necessarily zero.  From this
matrix, one finds, for example, that $3$-body flavor operators
transform as $1 + 35 + 405 + 2695$, but that pure $3$-body $2695$
operators only appear in $3$-body flavor operators.
\section{Mass Relations}
	We now study the mass relations which can be obtained using
the operator expansions of the previous section.  In Table~II, we
compile the mass combinations associated with the neglect of each of
the 19 operators of Sec.~III, and with the irreducible representations
of $SU(6)$ whenever the two do not coincide.  The mass combinations
are divided into isospin sectors $I=0,1,2,3$.  The definitions of the
baryon isospin mass combinations used in the table are presented in
the subsections which follow.  The mass combinations associated with
the 19 operators in Eqs.~(\ref{izero}--\ref{ithree}) are labeled by
their coefficients $c_{(n)}^{R,I}$.  The $1/\N$ suppressions and the
flavor-breaking parameters associated with each mass combination are
tabulated.  The $1/\N$ suppression factors assigned to each mass
combination include the implicit $\N$-dependence of operator matrix
elements as well as the explicit $1/\N$ factors displayed in
Eq.~(\ref{izero}--\ref{ithree}).  The flavor-breaking suppression
factors are obtained from Table~I.  Mass combinations which also
correspond to single $SU(6)$ representations appear with a check in
the column with heading label $SU(6)$; otherwise a ``No'' appears.
The $SU(6)$ mass combinations which differ from the above combinations
are listed in the subsequent blocks of the table.  These combinations
are labeled by coefficients $c^{R,I}_{D}$ where the subscript denotes
the dimension of the $SU(6)$ representation.  The $1/\N$ and the
leading flavor-breaking suppressions for the $SU(6)$ combinations are
listed.  Note that the flavor suppressions for the $SU(6)$
combinations follow from the perturbative flavor-breaking operator
analysis and are not consequences of the analysis in terms of
operators with definite $SU(6)$ transformation properties.

	We can understand which mass combinations in Table~II coincide
with $SU(6)$ mass combinations.  The mass combination associated with
the neglect of each operator in the basis
Eqs.~(\ref{izero}--\ref{ithree}) is broken by this operator alone,
{\it i.e.\/} all of the other operators in the basis vanish on this
mass combination.  Recall (\ref{matrix}) that the $3$-body flavor
operators are the only operators in the basis which contain components
transforming according to the $2695$ representation of $SU(6)$.  Since
all of the other $SU(6)$ representations occur in lower-body operators
as well, the only mass combination which vanishes for all lower-body
flavor operators are mass combinations in the $2695$.  Therefore, all
of the $3$-body mass combinations in the first block of Table~II
correspond to mass combinations in the $2695$ representation of
$SU(6)$.  In addition, for the $SU(3)$ singlet expansion $M^1_0$,
which involves only even-body operators, neglect of the 2-body
operator $J^2$ results in a mass combination in the $405$
representation of $SU(6)$.  In general, neglect of the highest-body
operator in any perturbative flavor symmetry-breaking expansion leads
to a mass combination corresponding to a definite $SU(6)$
representation.  None of the mass combinations following from the
neglect of other $n$-body operators appearing in
Eqs.~(\ref{izero}--\ref{ithree}) has this property.

	The last column of Table~II presents the experimental accuracy
of mass relations obtained by setting each mass combination equal to
zero.  These percentage accuracies are determined by evaluating the
given mass combination, and then dividing by one-half the sum of the
absolute values of all of the terms in the same mass combination.
Stated another way, the combination is organized as {\em lhs = rhs},
where {\em lhs} and {\em rhs} possess only baryon masses with positive
coefficients, and then we compute $|{(lhs-rhs)}/{((lhs+rhs)/2)}|$.
The purpose of this normalization is twofold: first, this expression
is invariant under multiplication of all baryon coefficients by the
same constant, and second, the resulting number is dimensionless.
Because the denominator is a sum of baryon masses with the same size
coefficients as in the numerator, the ratio represents a
scale-invariant measure of how much the numerator mass difference is
suppressed.  The experimental accuracies listed in Table~II are
obtained using the measured baryon masses and mass differences of the
Particle Data Group~\cite{pdg}, although there is controversy over
isospin splittings in the decuplet (see, {\it
e.g.\/}~\cite{lebeddec}).  The $\Delta^-$ mass is unmeasured; we
eliminate this mass from all of the $I=0,1,2$ $\Delta$ mass
combinations using the sole $I=3$ mass relation (see Subsec.~D), which
is satisfied to high orders in the $1/\N$ expansion and flavor
symmetry breaking.  Nevertheless, large experimental uncertainties
remain in the $I=1$ and $2$ isospin splittings of the $\Delta$, and
this prevents a meaningful comparison of the predicted theoretical
hierarchy of many of the $I=1$ and $2$ mass relations with experiment.

	The experimental accuracies in Table~II are to be compared
with the the combined $1/\N$ and flavor symmetry-breaking suppressions
of a given mass combination.  Note that the singlet baryon mass
proportional to the operator $\openone$ is order $\N$, so the $1/\N$
suppression of each mass combination relative to this singlet mass
contains one more power of $1/\N$ than what is listed in the table.
In the following four subsections, we analyze the mass relations
arising in the $I=0,1,2,3$ channels, respectively.

\subsection{$I=0$ Mass Relations}
	There are eight linear combinations of the octet and decuplet
masses which transform as $I=0$: $\Lambda$, $\Omega$, and
\begin{eqnarray}
N_0        & = & \frac 1 2 \left( p + n \right), \nonumber\\
\Sigma_0   & = & \frac 1 3 \left( \Sigma^+ + \Sigma^0 + \Sigma^-
\right), \nonumber\\
\Xi_0      & = & \frac 1 2 \left( \Xi^0 + \Xi^- \right), \nonumber\\
\Delta_0   & = & \frac 1 4 \left( \Delta^{++} +\Delta^+ +\Delta^0
+\Delta^- \right), \\
\Sigma^*_0 & = & \frac 1 3 \left( \Sigma^{*+} +\Sigma^{*0}
+\Sigma^{*-} \right), \nonumber\\
\Xi^*_0    & = & \frac 1 2 \left( \Xi^{*0} + \Xi^{*-} \right).
\nonumber
\end{eqnarray}

	The eight $I=0$ mass combinations in the first block of
Table~II correspond to the eight operators in Eq.~(\ref{izero}).  Each
mass combination is assigned definite $1/\N$ and $SU(3)$ flavor
symmetry-breaking suppression factors. The $SU(3)$ flavor
symmetry-breaking assignments are easy to understand: the octet
combinations arise at first order in $\epsilon$; the $27$ combinations
at order $\epsilon^2$; and the $64$ at order $\epsilon^3$.  The
experimental accuracies in the last column of Table~II allow us to
compare the relative suppressions of each operator in the $1/\N$
expansion; values are given for all of the mass combinations except
the singlet combination $c^{1,0}_{(0)}$, which is not
suppressed\footnote{The experimental accuracy for this mass
combination is $O(1)$, as expected.}.  The experimental accuracies
exhibit the hierarchy predicted by the combined $1/\N$ and flavor
symmetry-breaking suppressions; mass combinations which are more
highly suppressed in the $1/\N$ and $\epsilon$ expansions correspond
to proportionally smaller numbers in Table~II.  In addition,
comparison of the experimental accuracies of the mass combinations
$c^{1,0}_{(2)}$ and $c^{8,0}_{(1)}$ indicates that the flavor
symmetry-breaking parameter $\epsilon$ is indeed comparable to one
factor of $1/\N$.

	$I=0$ mass relations are obtained by successively neglecting
operators in the mass expansion.  The most highly suppressed operator
is the unique $64$ operator occurring at order $\epsilon^3/\N^2$ in
the combined $1/\N$ and flavor-breaking expansions.  Neglect of this
operator yields the mass relation
\begin{equation}\label{mrelone}
\frac 1 2 (\Delta_0 - 3 \Sigma^*_0 + 3 \Xi^*_0 - \Omega ) =0,
\end{equation}
with an expected relative accuracy of $\epsilon^3/\N^3$\footnote{This
relation is also satisfied by one-loop diagrams in chiral perturbation
theory\cite{jenkXPT}.}.  The naive estimate of the quantity
$\epsilon^3/\N^3$ is consistent with the experimental value $0.09 \pm
0.03 \%$.  At next subleading order, $O(\epsilon^2/N_c^2)$, the
operator $c^{27,0}_{(3)}$ is neglected, resulting in the mass relation
\begin{eqnarray}\label{zerotwo}
\lefteqn{2 \left(\frac{3}{4}\Lambda +\frac{1}{4}\Sigma_0 -\frac{1}{2}
(N_0 +\Xi_0)\right) =} && \nonumber\\
&&-\frac 1 7 (4\Delta_0 - 5 \Sigma^*_0 -2 \Xi^*_0 +3 \Omega ) .
\end{eqnarray}
Note that the left-hand side of Eq.~(\ref{zerotwo}) is proportional to
the Gell-Mann--Okubo formula, whereas the right-hand side is one
linear combination of the two decuplet equal spacing rules
(Eq.~(\ref{mrelone}) is the other).  The predicted $\epsilon^2/N_c^3$
accuracy of relation~(\ref{zerotwo}) is consistent with the
experimental value.  To next order in the combined $1/\N$ and
$\epsilon$ expansions, one neglects the two mass operators
$c^{8,0}_{(3)}$ and $c^{27,0}_{(2)}$ occurring at the comparable
orders $O(\epsilon/N_c^2)$ and $O(\epsilon^2/N_c)$, respectively.  The
two mass relations listed in Table~II are satisfied at the expected
level of accuracy.  Note that mass relation $c^{27,0}_{(2)}$ is
satisfied to a somewhat greater level of accuracy than mass relation
$c^{8,0}_{(3)}$, although both are consistent with the canonical
values from the combined $\epsilon$ and $1/\N$ expansions with
natural-size coefficients.  By neglecting the $c^{8,0}_{(2)}$
operator, one obtains another mass relation at $O(\epsilon/\N)$, with
canonical accuracy of $3\%$ as compared to an experimental value of
$6\%$.  The two final mass combinations $c^{8,0}_{(1)}$ and
$c^{1,0}_{(2)}$ arise at the comparable orders $1/\N$ and $\epsilon$,
respectively, so neither operator can be neglected relative to the
other, and no additional mass relation is obtained.  It is valid,
however, to neglect the $c^{1,0}_{(2)}$ operator relative to the
$c^{1,0}_{(0)}$ operator when considering the $1/\N$ expansion in the
singlet channel.  The mass relation following from the neglect of the
$c^{1,0}_{(2)}$ operator $J^2/\N$ in the singlet mass expansion is the
trivial relation which equates the average octet and decuplet masses,
\begin{eqnarray}
\frac 1 8 ( 2 N_0 + 3 \Sigma_0 &&+ \Lambda + 2\Xi_0 ) = \nonumber\\
&&\frac 1 {10} ( 4 \Delta_0 + 3 \Sigma^*_0 + 2 \Xi^*_0 + \Omega ) .
\end{eqnarray}
The predicted accuracy of this relation, which is a measure of the
size of decuplet-octet mass splittings, is of order
\begin{equation}
\frac{1}{N_c^2} \left( (J^2)_{3/2} - (J^2)_{1/2} \right) =
\frac{1}{N_c^2} \left(\frac{15}{4} -\frac{3}{4}\right) =
\frac{3}{\N^2} ,
\end{equation}
which is consistent with the experimental value of $18\%$.

	Finally, we consider how our results relate to previous
analyses.  It is possible to analyze the $I=0$ baryon masses in a
$1/\N$ expansion alone, in contrast to the combined $1/\N$ and flavor
symmetry-breaking expansion of this work.  In the $1/\N$ expansion,
the most highly suppressed $I=0$ operators are $O(1/N_c^2)$, namely
the operators with coefficients $c^{8,0}_{(3)}$, $c^{27,0}_{(3)}$ and
$c^{64,0}_{(3)}$.  Neglect of these operators yields the three mass
relations in Table~II corresponding to these three coefficients.  An
equivalent set of mass relations is
\begin{eqnarray}\label{newthree}
& & \Sigma^*_0 - \Sigma_0 = \Xi^*_0 -\Xi_0 \hspace{2em}
(0.88 \pm 0.02\%), \nonumber\\ & &
\frac{3}{4}\Lambda +\frac{1}{4}\Sigma_0 -\frac{1}{2} (N_0 +\Xi_0)
\nonumber \\
& & \hspace{2em} = -\frac{1}{4} (\Omega -\Sigma^*_0 -\Xi^*_0 +
\Delta_0) \hspace{1.7em} (0.16 \pm 0.01\%), \\
& & {1 \over 2}\left(\Sigma_0^* -
\Delta_0 \right) -
\left(\Xi^*_0 - \Sigma^*_0
\right) + {1 \over 2}\left(\Omega - \Xi^*_0 \right)=0, \nonumber
\end{eqnarray}
where the first relation is a linear combination of the
$c^{8,0}_{(3)}$, $c^{27,0}_{(3)}$ and $c^{64,0}_{(3)}$ mass relations;
the second is a linear combination of the $c^{27,0}_{(3)}$ and
$c^{64,0}_{(3)}$ relations; and the third is the $c^{64,0}_{(3)}$ mass
relation~(\ref{mrelone}).  Eq.~(\ref{newthree}) is the set of mass
relations obtained in Refs.~\cite{djm,djmtwo} using $SU(2) \times
U(1)$ symmetry.  Since the $c^{8,0}_{(3)}$, $c^{27,0}_{(3)}$ and
$c^{64,0}_{(3)}$ operators are the $I=0$ operators associated with the
$2695$ of $SU(6)$, these relations also were obtained in the $SU(6)$
$I=0$ analysis of Ref.~\cite{lebed}.  The experimental accuracies of
these relations clearly exhibit the hierarchy
$\epsilon/\N^2$:$\epsilon^2/\N^2$:$\epsilon^3/\N^2$, so there is
evidence for both the $1/\N$ and flavor-breaking suppressions.  In a
second analysis, Ref.~\cite{djmtwo} obtained $I=0$ mass relations
valid at linear order in the flavor symmetry breaking $\epsilon$ and
neglecting operators suppressed by $1/\N^2$.  This analysis
corresponds to the neglect of the $c^{8,0}_{(3)}$, $c^{27,0}_{(2)}$,
$c^{27,0}_{(3)}$ and $c^{64,0}_{(3)}$ operators, all of which are
suppressed by $1/\N^2$ and/or $\epsilon^2$.  The four resulting mass
relations (see Eqs.~(10.11)-(10.14) of Ref.~\cite{djmtwo}) are the
Gell-Mann--Okubo formula, the two decuplet equal spacing rules, and
$\Sigma^*_0 -\Sigma_0 = \Xi^*_0 -\Xi_0 $.  In our analysis, the same
set is obtained by truncating at order $\epsilon/\N^2 \sim
\epsilon^2/\N$ in the combined $1/\N$ and flavor symmetry-breaking
expansions.  The analysis of this work, however, exhibits the complete
$1/\N$ and symmetry-breaking structure of these relations, and obtains
the one additional relation at order $\epsilon/\N$.  From our
analysis, we are able to conclude that the $1/\N$ and flavor
symmetry-breaking suppression factors are both required to describe
the observed hierarchy of the $I=0$ sector.
\subsection{$I=1$ Mass Relations}
	There are seven $I=1$ mass combinations: six $I=1$ mass
splittings
\begin{eqnarray}
N_1        & = & (p-n), \nonumber\\
\Sigma_1   & = & (\Sigma^+ - \Sigma^-), \nonumber\\
\Xi_1      & = & (\Xi^0 - \Xi^-), \nonumber\\
\Delta_1   & = & (3 \Delta^{++} + \Delta^+ - \Delta^0 - 3 \Delta^-),
\\
\Sigma^*_1 & = & (\Sigma^{*+} - \Sigma^{*-}), \nonumber\\
\Xi^*_1    & = & (\Xi^{*0} - \Xi^{*-}), \nonumber
\end{eqnarray}
and one off-diagonal mass $\Lambda\Sigma^0$.

	The most highly suppressed $I=1$ operator in the combined
$1/\N$ and flavor-breaking expansions is the unique $64$ operator
occurring at order $\epsilon^\prime \epsilon^2/\N^2$.  Neglect of this
operator yields the mass relation
\begin{equation}\label{ionemassrel}
\Delta_1 - 10 \Sigma^*_1 + 10 \Xi^*_1 =0 ,
\end{equation}
with a predicted accuracy of $\epsilon^\prime \epsilon^2/\N^3 \approx
10^{-5}$.  A meaningful comparison of this suppression factor with
experiment is not possible at present because of large experimental
uncertainties in decuplet $I = 1$ splittings, particularly $\Delta_1$.
At next subleading order in the combined $1/\N$ and flavor
symmetry-breaking expansions, one neglects the four operators
$c^{8,1}_{(3)}$, $c^{27,1}_{(2)}$, $c^{27,1}_{(3)}$ and
$c^{10+\overline{10},1}_{(3)}$ occurring at comparable orders
$\epsilon^\prime/\N^2$ and $\epsilon^\prime \epsilon/\N$.  The
$c^{10+\overline{10},1}_{(3)}$ mass relation is the Coleman--Glashow
relation
\begin{equation} \label{cg}
N_1 - \Sigma_1 + \Xi_1 =0,
\end{equation}
which is known to be very accurate; the experimental accuracy for this
relation is consistent with zero.  The central value, however, is
completely in line with the naive estimate of the quantity
$\epsilon^\prime \epsilon/ \N^2$, where $\epsilon^\prime$ is
numerically about $1/\N^5$ in QCD.  More accurate measurements of
$\Sigma_1$ and $\Xi_1$ are required to fully test this relation at the
level predicted by our analysis.  Notice that the two combinations
$c^{27,1}_{(2)}$ and $c^{27,1}_{(3)}$ are exactly the same order in
$1/\N$ and flavor breaking, and exhibit the same combinations of octet
and decuplet masses.  Thus, an equivalent, simpler pair of mass
relations at order $\epsilon^\prime \epsilon/\N$ is
\begin{eqnarray}
N_1 -\Xi_1 +2\sqrt{3}\Lambda\Sigma^0 & = & 0, \label{lamsig} \\
\Delta_1 -3\Sigma^*_1 -4\Xi^*_1 & = & 0. \label{c272}
\end{eqnarray}
Relation~(\ref{c272}) is satisfied to an experimental accuracy of
$0.27 \pm 0.10\%$.  Again, the large uncertainty in $\Delta_1$
prevents a meaningful comparison of this value with the theoretical
accuracy of $\epsilon^\prime \epsilon/\N^2$.  The other two relations
at this order in the expansion, $c^{8,1}_{(3)}$ and
Eq.~(\ref{lamsig}), involve the unmeasured $\Lambda \Sigma^0$ mass,
and cannot be compared with experiment.  Finally, the two remaining
mass combinations $c_{(1)}^{8,1}$ and $c_{(2)}^{8,1}$ both appear at
order $\epsilon^\prime$ in the combined $1/\N$ and flavor
symmetry-breaking expansions, so neither operator can be neglected
relative to the other, and no additional mass relation is obtained.

	Comparison of our $I=1$ mass hierarchy with experiment is
limited by the large experimental uncertainty in the splitting
$\Delta_1$ and the presence of the unknown parameter $\Lambda
\Sigma^0$.  It is possible to extract additional information about the
$I=1$ mass hierarchy by eliminating these uncertain parameters.  One
may add to a given mass combination any other combinations which are
of the same or higher order in the combined flavor and $1/\N$
expansions.  Such a linear combination remains at the same order in
the combined expansions as the original one, although it no longer
necessarily corresponds to a single $SU(3)$ representation.  We use
the $c^{64,1}_{(3)}$ mass relation Eq.~(\ref{ionemassrel}) to
eliminate $\Delta_1$ from all other $I=1$ mass combinations, and
Eq.~(\ref{lamsig}) to eliminate $\Lambda\Sigma^0$.  In its own right,
this equation predicts
\begin{equation}
\Lambda\Sigma^0 = -1.47 \pm 0.17 \mbox{ MeV}.
\end{equation}
We now analyze the results of these substitutions.  The
$c^{10+\overline{10},1}_{(3)}$ mass combination does not involve
either $\Delta_1$ or $\Lambda \Sigma^0$, and so is unaffected by this
procedure.  With these substitutions, the four remaining mass
combinations $c^{8,1}_{(1)}$, $c^{8,1}_{(2)}$, $c^{8,1}_{(3)}$ and
Eq.~(\ref{c272}) reduce respectively to the four mass combinations
given in the third block of the $I=1$ sector of Table~II.  Note that
the $1/\N$ and flavor symmetry-breaking assignments of these
combinations are identical to those of the original $SU(3)$
combinations.  The assignments for the third combination do not appear
in the table; it is the linear combination of $c^{8,1}_{(3)}$ and
Eq.~(\ref{lamsig}) which eliminates $\Lambda
\Sigma^0$, and so combines order $\epsilon^\prime \epsilon/\N$ and
$\epsilon^\prime/\N^2$ contributions, which are comparable in the
combined $1/\N$ and flavor-breaking expansions.

	From experimental values for these four mass combinations, we
conclude that our predicted flavor-breaking and $1/\N$ hierarchy is
also evident in the $I=1$ splittings.  The first two combinations are
expected to work at the level $\epsilon^\prime /\N$, and their
experimental accuracies are similar and consistent with
$\epsilon^\prime \approx 1/\N^5$.  The last relation
\begin{equation}
\Sigma^*_1 = 2 \Xi^*_1
\end{equation}
has an expected accuracy of $\epsilon^\prime \epsilon/\N^2$.  The
central value of the experimental accuracy is consistent with a
suppression of $\epsilon/\N$ relative to the first two mass
combinations.  The error on this experimental accuracy is large,
however.  The central value of the experimental accuracy of the third
relation is surprisingly small, but its uncertainty puts it into the
same range as that of the fourth relation.  Likewise, the central
value of the experimental accuracy for the
$c^{10+\overline{10},1}_{(3)}$ relation lies within this same range.
However, none of these last three relations is measured accurately
enough to test our predicted hierarchy conclusively.  Reasonable
improvements in the measurement of $I=1$ mass splittings would enable
a substantive comparison.
\subsection{$I=2$ Mass Relations}
	There are three $I=2$ splittings:
\begin{eqnarray}
\Sigma_2   & = & (\Sigma^+ - 2 \Sigma^0 + \Sigma^-), \nonumber\\
\Delta_2   & = & (\Delta^{++} - \Delta^+ - \Delta^0 + \Delta^-), \\
\Sigma^*_2 & = & (\Sigma^{*+} - 2 \Sigma^{*0} + \Sigma^{*-}).
\nonumber
\end{eqnarray}

	The most highly suppressed $I=2$ operator in the combined
$1/\N$ and flavor-breaking expansions is the unique $64$ operator
occurring at order $\epsilon^{\prime\prime} \epsilon/\N^2$.  Neglect
of this operator yields the mass relation
\begin{equation}
\Delta_2 = 2\Sigma^*_2 ,
\end{equation}
with a predicted accuracy of $\epsilon^{\prime\prime} \epsilon/\N^3
\approx 3 \times 10^{-5}$.  A meaningful comparison of this
suppression factor with experiment is not possible because of large
experimental uncertainties in the decuplet $I=2$ splittings.  The two
remaining mass combinations $c^{27,2}_{(2)}$ and $c^{27,2}_{(3)}$ are
both order $\epsilon^{\prime\prime} \epsilon/\N$, so neither operator
can be neglected relative to the other, and no additional mass
relation is obtained.

	The hierarchy of $I=2$ mass combinations is completely
consistent with the predictions of the combined $1/\N$ and
flavor-breaking expansions.  Recall that the $I=2$ flavor
symmetry-breaking parameter $\epsilon^{\prime\prime}$ is comparable to
the $I=1$ parameter $\epsilon^\prime$.  Notice from Table~II that all
$I=2$ combinations (and hence any linear combination of them) are
suppressed by one factor of $1/\N$ relative to the largest $I=1$
combinations.  The experimental accuracy of the $c^{27,2}_{(2)}$
combination is suppressed at this level relative to the two measured
$O(\epsilon^\prime)$ $I=1$ mass combinations in Table~II.  The
$c^{27,2}_{(3)}$ combination may also be suppressed at this level, but
its experimental accuracy is too poorly known.  In addition, these two
$I=2$ combinations are predicted to be suppressed by
$\epsilon^{\prime\prime} \approx 5 \times 10^{-3}$ relative to the
largest $I=0$ mass combination $c^{1,0}_2$, and consistency with this
prediction is also borne out in Table~II.  Unfortunately, however, the
uncertainties of the experimental accuracies for the $I=2$ mass
combinations are substantial compared to their central values, so one
cannot draw definitive conclusions about the accuracy of the observed
hierarchy from the $I=2$ masses.
\subsection{$I=3$ Mass Relations}
	The lowest-lying baryons admit only one $I=3$ splitting,
\begin{equation}
\Delta_3 = (\Delta^{++} -3\Delta^+ +3\Delta^0 -\Delta^-).
\end{equation}
This mass combination corresponds to the single $I=3$ operator, with
coefficient $c_{(3)}^{64,3}$, that arises at order
$\epsilon^{\prime\prime} \epsilon^\prime /N_c^2$ in the combined
$1/\N$ and flavor-breaking expansions.  Neglecting the
$c_{(3)}^{64,3}$ operator yields the mass relation
\begin{equation}\label{ithreerel}
\Delta_3 = 0.
\end{equation}
The suppression factor $\epsilon^{\prime\prime} \epsilon^\prime
/N_c^3$ is second order in the isospin-breaking parameters, and so is
much smaller than any of the other suppression factor in our analysis.
Thus, we expect violations of Eq.~(\ref{ithreerel}) to be quite small.
A naive estimate of the size of $\Delta_3$ gives of order
$10^{-3}$~MeV at most.  A calculation~\cite{lebeddec} of this quantity
in chiral perturbation theory, including loop effects, does not alter
this conclusion.  We have used the extreme accuracy of the mass
relation~(\ref{ithreerel}) to eliminate the unknown $\Delta^-$ mass in
the $I=0,1,2$ $\Delta$ mass splittings.
\section{Completely Broken $SU(3)$ Symmetry}
	The analysis of the $I=0,1,2,3$ baryon mass splittings can be
performed using only $SU(2) \times U(1)$ flavor symmetry.  Such an
analysis yields mass relations which are valid to all orders in
$SU(3)$ symmetry breaking.  In this section, we reanalyze the baryon
isospin mass splittings using $SU(2) \times U(1)$ flavor symmetry,
treating isospin breaking as a small perturbation.  The relevant
spin-flavor symmetry group is $SU(4) \times SU(2) \times U(1)$, where
the $SU(4)$ factor is the spin-flavor group of the two light flavors
$u$ and $d$; the $SU(2)$ factor is strange quark spin; and the $U(1)$
factor is the number of strange quarks.  The analysis of the $I=0$
sector was performed in Refs.~\cite{djm,djmtwo}, so we restrict the
analysis here to $I=1,2,3$ mass combinations.

	The $SU(2) \times U(1)$ operator analysis uses one-body
operators with definite isospin and strangeness instead of operators
with definite $SU(3)$ transformation properties.  In particular, this
implies that the one-body operators $J_{ud}^i = J_u^i + J_d^i$ and
$J_s^i$ are used instead of $J^i$ and $G^{i8}$, and that the strange
quark number operator $N_s$ is used instead of $T^8$.  An $I \neq 0$
operator has the generic form
\begin{equation}\label{iops}
\N \left( \frac{I^3}{\N} \right)^p \left(\frac{1}{N_c^2} \{
J_s^i , G^{i3} \} \right)^q ,
\end{equation}
where $I=p+q$, times polynomials in ${N_s}/{\N}$, ${I^2}/{N_c^2}$ and
${J^2}/{N_c^2}$ (see Ref.~\cite{djmtwo}).  In analogy with Sec.~II, we
will restrict our analysis to the eleven isospin mass splittings of
the physical baryons.  For $\N>3$, we identify the physical baryon
states with states at the top of the weight diagrams Figs.~3 and~4
that have strangeness, isospin, and spins $J$, $J_{ud}$ and $J_s$ of
order one.  The operator $\{ J_s^i, G^{i3} \}$ has non-trivial matrix
elements of order $\N$ for the physical baryons, so the matrix
elements of $I^3$ and $\{ J_s^i, G^{i3} \}/\N$ are both $O(1)$ for
these states.  When we restrict our set of baryon states to the
physical baryons, we only need to retain operators up to $3$-body
operators; $4$- or higher-body operators are either redundant or
vanish on this set.  For $\N>3$, we find a total of eleven independent
$0$-, $1$-, $2$-, and $3$-body operators.  These same eleven operators
also result from immediate specialization to $\N =3$.

	The eleven $I \not= 0$ operators which span the $I=1,2,3$
isospin splittings of the baryon octet and decuplet are generated
using Eq.~(\ref{iops}) and truncating at $3$-body operators.  In the
following, we treat isospin symmetry breaking perturbatively.  Because
the isospin-breaking parameters $\epsilon^\prime$ and
$\epsilon^{\prime\prime}$ are both comparable to effects of order
$1/\N^5$, it is sufficient to work to linear order in these
parameters.  There are two operators
\begin{equation} \label{lead}
c^{\,1}_{(1)} I^3 + c^{\,1}_{(2)} \frac{1}{\N} \{ J_s^i, G^{i3} \} ,
\end{equation}
at leading order, $O(\epsilon^\prime)$, in the combined $1/\N$ and
isospin-breaking expansions, and four operators at next-to-leading
order, $O(\epsilon^\prime/\N)$ or $O(\epsilon^{\prime\prime}/\N)$,
\begin{eqnarray} \label{nlo}
& & d^{\,1}_{(2)} \frac{N_s}{\N} I^3 + d^{\,1}_{(3)} \frac{N_s}{N_c^2}
\{ J_s^i, G^{i3} \} \nonumber \\
& & \mbox{  } + d^{\,2}_{(2)} \frac{1}{\N} \{ I^3, I^3 \} +
d^{\,2}_{(3)} \frac{1}{N_c^2} \{ I^3, \{ J_s^i, G^{i3} \} \} .
\end{eqnarray}
The operators in Eqs.~(\ref{lead}) and~(\ref{nlo}) are understood to
possess a single value of isospin as indicated by the coefficient
superscripts; the subtraction of smaller isospin representations than
the one indicated is implicit.  The $I=2$ operators appearing at
next-to-leading order originate in electromagnetic interactions, as
discussed in Sec.~III, with comparable coefficients
($O(\epsilon^{\prime\prime})$) to those of $I=1$ operators
($O(\epsilon^\prime)$).  Note that there is one additional operator
which must be included for $\N>3$ if one does not restrict to physical
baryons, namely the $4$-body operator
\begin{equation}
\frac{1}{N_c^3} \{ \{ J_s^i, G^{i3} \} , \{J_s^i, G^{i3} \} \} .
\end{equation}
Finally, there are four additional $3$-body operators at
$O(\epsilon^\prime/\N^2)$ or $O(\epsilon^{\prime\prime}/\N^2)$, $I^2
I^3$, $J^2 I^3$, $N_s^2 I^3$ and $N_s \{ I^3, I^3 \}$, and one
additional $3$-body operator $\{ I^3, \{ I^3, I^3 \}\}$ at
$O(\epsilon^{\prime\prime}\epsilon^\prime/\N^2)$.  This last operator
is second order in isospin breaking parameters.  Note that there are
additional higher-body operators at these orders which must be
included if the set of baryon states is not restricted to the physical
baryons.

	Mass relations are obtained by successive neglect of $1/\N$
suppressed operators.  At leading order in the combined $1/\N$ and
isospin breaking expansions, one retains the two operators
Eq.~(\ref{lead}), which implies five mass relations amongst the seven
$I=1$ combinations.  One obtains this same number of mass relations in
the perturbative $SU(3)$ analysis of Sec.~IV, but the spaces spanned
by these two sets of relations are not equivalent, because the
operator bases are not exactly the same.  Specifically,
\begin{equation}
\{ J^i, G^{i3} \} = \{ J_{ud}^i, G^{i3} \} + \{ J_s^i , G^{i3} \},
\end{equation}
where
\begin{equation}
\{ J_{ud}^i , G^{i3} \} = \frac{1}{2} (\N -N_s +2) I^3
\end{equation}
by the operator identities\cite{djmtwo}, so the $SU(2) \times U(1)$
case introduces a higher-order piece (the $N_s I^3$) than present at
the same order in the $SU(3)$ case.  Nevertheless, four of the five
relations coincide; they may be written as the combinations
$c^{8,1}_{(3)}$, $c^{27,1}_{(3)}$, $c^{10+\overline{10},1}_{(3)}$ and
$c^{64,1}_{(3)}$ from Table~II (the combination $c^{27,1}_{(2)}$ is
broken by Eq.~(\ref{lead})).  However, there is no reason to single
out combinations corresponding to unique $SU(3)$ representations in
the completely broken $SU(3)$ analysis.  We instead choose linearly
independent combinations with the smallest possible experimental
uncertainties, so that one obtains the most stringent test of the
$1/\N$ hierarchy.  From Eq.~(\ref{nlo}), one sees that two new $I=1$
operators appear at next-to-leading order.  Thus, two of the five
$O(1)$ relations are broken at $O(1/\N)$; we choose them to be the
Coleman--Glashow relation (\ref{cg}) and the combination
\begin{equation} \label{brk2}
3N_1 -\Xi_1 -2\Xi^*_1 =0 \hspace{2em} (0.12 \pm 0.02\%),
\end{equation}
where the number given is the experimental accuracy of the relation as
defined in Sec.~IV.  Furthermore, since the two leading-order
operators (\ref{lead}) are $I=1$, no $I=2$ or $I=3$ splittings are
produced at $O(1)$ in the $1/\N$ expansion.

	At next-to-leading order (\ref{nlo}), two new $I=1$ operators
appear, so we expect three $I=1$ relations to remain.  This counting
seems to be at odds with the $1/\N$ factors given in Table~II for the
perturbative $SU(3)$ case.  In the $SU(3)$ analysis, one forms
symmetric and antisymmetric combinations of an $O(1/\N)$ and an
$O(1/\N^2)$ operator to obtain the pure $SU(3)$ $27$ and
$10+\overline{10}$ representations, so that one finds instead two
$O(1/\N)$ combinations.  In the present analysis, these two operators
remain unmixed.  We find that none of the combinations associated with
the pure $SU(3)$ representations survive at this order (because
$SU(3)$ is completely broken, this result is perhaps not surprising).
A convenient choice for the set of three $I=1$ relations at $O(1/\N)$
is
\begin{eqnarray}
10N_1  =  \Delta_1 \hspace{2em} (0.19 \pm 0.10\%), && \label{brk3} \\
-3N_1+(\Sigma_1 -\Xi_1) +2(\Sigma^*_1 - \Xi^*_1) = 0 && \nonumber \\
(0.002 \pm 0.016\%), && \label{brk4} \\
2(\Sigma_1 - \Sigma^*_1) - (\Xi_1 - \Xi^*_1) = 2\sqrt{3} \Lambda
\Sigma^0 . && \label{brk5}
\end{eqnarray}
The final relation predicts
\begin{equation}
\Lambda \Sigma^0 = -1.20 \pm 0.43 \mbox{ MeV}.
\end{equation}
In addition to the $I=1$ operators, two $I=2$ operators appear at
$O(1/\N)$, so there is one $I=2$ relation at this order,
\begin{equation}
2\Sigma_2 -3\Delta_2 + 4\Sigma^*_2 = 0 \hspace{2em} (0.20 \pm 0.08\%),
\end{equation}
and one $I=3$ relation (\ref{ithreerel}).  All of these relations are
violated at order $1/\N^2$; the $I=3$ relation also requires an
additional factor of isospin breaking.

	The immediate conclusion we obtain from the analysis of this
section is that the $SU(2) \times U(1)$ analysis does not produce as
good a hierarchy as the perturbative $SU(3)$ analysis of the previous
section.  If we believe only in completely broken $SU(3)$, then the
$O(1)$ $I=1$ relations Eqs.~(\ref{cg}) and (\ref{brk2}) should display
accuracies of $O(\epsilon^\prime /\N) \approx 0.15\%$, and the
$O(1/\N)$ $I=1$ relations Eqs.~(\ref{brk3})--(\ref{brk5}) accuracies
of $O(\epsilon^\prime /N_c^2) \approx 0.05\%$.  In the $SU(2) \times
U(1)$ analysis, we cannot explain why Eqs.~(\ref{cg}) and (\ref{brk4})
work so well experimentally.  This fact provides additional
evidence for
the perturbative $SU(3)$ flavor-breaking analysis of Sec.~IV: not only
are the perturbative results consistent with experiment, but the
accuracy of some mass relations cannot be explained otherwise.  This
conclusion is most obvious in the $I=0$ sector; the analysis of this
section shows that there is also evidence for it in the $I =1$ sector.
\section{Conclusions}
	In summary, we conclude that there is striking evidence for
the mass hierarchy predicted by a combined expansion in $1/\N$ and
$SU(3)$ flavor symmetry breaking, with flavor breaking treated
perturbatively.  Neither a $1/\N$ nor a flavor expansion alone
explains the observed hierarchy.  In addition, a $1/\N$ expansion
treating only isospin breaking perturbatively fails to explain the
hierarchy of the $I=0$ and $I=1$ mass combinations, so it is clearly
better to treat $SU(3)$ as a perturbatively broken, rather than
completely broken, symmetry.

	Our analysis explicitly shows that the combined expansion
differs from the old non-relativistic $SU(6)$ analysis, which
neglected only mass combinations in the $2695$.  In the $1/\N$
expansion, $2695$ combinations are usually suppressed by a factor of
$1/\N^2$, which accounts for the fact that many of the $1/\N$ mass
relations coincide with $SU(6)$ combinations.  There are additional
relations obtained in the $1/\N$ expansion satisfied at this same
level of accuracy, however, which are not members of the $2695$ and
are therefore missed in the old $SU(6)$ analysis.

	Finally, it is important to emphasize that improved
measurements of baryon mass splittings (particularly decuplet isospin
splittings) are needed to test a number of our mass relations at the
level of accuracy predicted by the combined expansion.  Even a modest
decrease of experimental uncertainties in some mass combinations would
be enough to permit one to distinguish conclusively between the
predictions of this method and those of other possible hierarchies.
\section*{Acknowledgments}
	This work was supported in part by the Department of Energy
under grant DOE-FG03-90ER40546.  E.J. was supported in part by NYI
award PHY-9457911 from the National Science Foundation.

\onecolumn 

\widetext

\begin{table}[htbp]
\caption{Flavor symmetry breaking of $M^R_I$}
\label{tab:one}
\smallskip
\centerline{\vbox{ \tabskip=0pt \offinterlineskip
\def\tablerule{\noalign{\hrule}}
\def\space{height 2pt&\omit&&\omit&&\omit&&\omit
&&\omit&&\omit&&\omit&&\omit&&\omit&&\omit&&\omit&\cr}
\halign{
\vrule #&\strut\hfil\ $ # $\ \hfil&&
\vrule #&\strut\hfil\ $ # $\ \hfil\cr
\tablerule\space
& M^1_0 && M^8_0 && M^{27}_0 && M^{64}_0 &&
M^{8}_1 && M^{27}_1 && M^{10 + \overline{10}}_1
&& M^{64}_1 && M^{27}_2 && M^{64}_2 &&
M^{64}_3 &\cr
\space\tablerule\space\space\space
& 1 && \epsilon && \epsilon^2 && \epsilon^3 &&
\epsilon^\prime && \epsilon^\prime \epsilon &&
\epsilon^\prime \epsilon &&
\epsilon^\prime \epsilon^2 &&
\epsilon^{\prime\prime} &&
\epsilon^{\prime\prime} \epsilon &&
\epsilon^{\prime\prime} \epsilon^\prime &\cr
\space\tablerule
}}}
\end{table}

\vfill\break\eject

\begin{table}[htbp]
\caption{Mass Combinations:  The isospin mass combinations appearing
in the table are defined in Sec.~IV.  Orders of $1/\N$ and flavor
symmetry breaking are given for each combination.  Experimental
accuracies appear in the final column.}
\smallskip
\label{tab:two}
\centerline{\vbox{ \tabskip=0pt \offinterlineskip
\def\tablerule{\noalign{\hrule}}
\def\space{height 2pt&\omit&&\omit&&\omit&&\omit&&\omit&&\omit&&
\omit\cr}
\halign{
\vrule #&\strut\hfil\ #\hfil&&
\vrule #&\strut\hfil\ $\rm #$\ \hfil\cr
\tablerule\space\tablerule
& && \omit \hfil $I=0$ \hfil && 1/\N && Flavor && SU(6) && Expt.
&&\omit\cr
\tablerule\space\tablerule\space
& $c^{1,0}_{(0)}$ && 25(2N_0 +3\Sigma_0
+\Lambda +2\Xi_0) -4(4\Delta_0 +3\Sigma^*_0 +2\Xi^*_0 +\Omega) && \N
&& 1 && No && * &&\omit\cr\space
& $c^{1,0}_{(2)}$ && 5(2N_0+ 3\Sigma_0 +\Lambda
+2\Xi_0) -4(4\Delta_0 +3\Sigma^*_0 +2\Xi^*_0 +\Omega) && 1/\N && 1
&&\surd && 18.21 \pm 0.03\% &&\omit\cr\space
& $c^{8,0}_{(1)}$ && 5(6N_0 -3\Sigma_0 +\Lambda
-4\Xi_0) -2(2\Delta_0 -\Xi^*_0 -\Omega) && 1 && \epsilon && No &&
20.21 \pm 0.02\% &&\omit\cr\space
& $c^{8,0}_{(2)}$ && N_0 -3\Sigma_0 +\Lambda
+\Xi_0 && 1/\N && \epsilon && No && 5.94 \pm 0.01\% &&\omit\cr\space
& $c^{8,0}_{(3)}$ && (-2N_0 -9\Sigma_0
+3\Lambda + 8\Xi_0) +2(2\Delta_0 -\Xi^*_0 -\Omega) && 1/\N^2 &&
\epsilon &&\surd && 1.11 \pm 0.02\% &&\omit\cr\space
& $c^{27,0}_{(2)}$ && 35(2N_0 -\Sigma_0
-3\Lambda +2\Xi_0) -4(4\Delta_0 -5\Sigma^*_0 -2\Xi^*_0 +3\Omega) &&
1/\N && \epsilon^2 && No && 0.37 \pm 0.01\% &&\omit\cr\space
& $c^{27,0}_{(3)}$ && 7 (2N_0 -\Sigma_0
-3\Lambda + 2\Xi_0) -2(4\Delta_0 -5\Sigma^*_0 -2\Xi^*_0 +3\Omega) &&
1/\N^2 && \epsilon^2 &&\surd && 0.17 \pm 0.02\% &&\omit\cr\space
& $c^{64,0}_{(3)}$ && \Delta_0 - 3 \Sigma^*_0 + 3
\Xi^*_0 - \Omega && 1/\N^2 && \epsilon^3 &&\surd && 0.09 \pm 0.03\%
&&\omit\cr\space
\tablerule\space
& $c^{1,0}_{1}$ && (2N_0 +3\Sigma_0 +\Lambda
+2\Xi_0) +2(4\Delta_0 +3\Sigma^*_0 +2\Xi^*_0 +\Omega) && \N && 1 &&
&& * &&\omit\cr\space
& $c^{8,0}_{35}$ && (N_0 -\Xi_0) +2(2\Delta_0 -\Xi^*_0 -\Omega) &&  1
&& \epsilon && && 27.44 \pm 0.04\% &&\omit\cr\space
& $c^{8,0}_{405}$ && (7N_0 -6\Sigma_0 +2\Lambda -3\Xi_0)
-2(2\Delta_0 -\Xi^*_0 -\Omega) && 1/\N && \epsilon && && 5.27 \pm
0.02\% &&\omit\cr\space
& $c^{27,0}_{405}$ && (2N_0 -\Sigma_0 -3\Lambda +2\Xi_0)
+ 4(4\Delta_0 -5\Sigma^*_0 -2\Xi^*_0 +3\Omega) && 1/\N && \epsilon^2
&& && 0.48 \pm 0.03\% &&\omit\cr\space
\tablerule\space\tablerule\space
&    && \omit \hfil $I=1$ \hfil &&   &&  &&
&&\omit&&\omit\cr
\tablerule\space\tablerule\space
& $c^{8,1}_{(1)}$ && 5(N_1 +5\Sigma_1 +4\Xi_1
+2\sqrt{3} \,\Lambda\Sigma^0) - (\Delta_1 +2\Sigma^*_1 +\Xi^*_1) && 1
&& \epsilon^\prime && No && \mbox{---} &&\omit\cr\space &
$c^{8,1}_{(2)}$ &&  -3N_1 +3\Xi_1 +4\sqrt{3}\,
\Lambda\Sigma^0 && 1 && \epsilon^\prime && No && \mbox{---}
&&\omit\cr\space
& $c^{8,1}_{(3)}$ && (-7N_1 -5\Sigma_1 +2\Xi_1 +
6\sqrt{3} \,\Lambda \Sigma^0) + (\Delta_1 +2\Sigma^*_1 +\Xi^*_1) &&
1/\N^2 && \epsilon^\prime &&\surd && \mbox{---} &&\omit\cr\space
& $c^{27,1}_{(2)}$ && 35(N_1 -\Xi_1 +2\sqrt{3}\,
\Lambda\Sigma^0) -2(\Delta_1 -3\Sigma^*_1 -4\Xi^*_1) && 1/\N &&
\epsilon^\prime\epsilon && No && \mbox{---} &&\omit\cr\space
& $c^{27,1}_{(3)}$ && 7(N_1 -\Xi_1
+2\sqrt{3} \,\Lambda \Sigma^0) -(\Delta_1 -3\Sigma^*_1 -4\Xi^*_1) &&
1/\N && \epsilon^\prime\epsilon &&\surd && \mbox{---} &&\omit\cr
\space
& $c^{10 + \overline{10},1}_{(3)}$ && N_1
-\Sigma_1 +\Xi_1 && 1/\N && \epsilon^\prime\epsilon &&\surd && 0.01
\pm 0.02\% &&\omit\cr\space
& $c^{64,1}_{(3)}$ && \Delta_1 - 10\Sigma^*_1
+10\Xi^*_1 && 1/\N^2 && \epsilon^\prime\epsilon^2 &&\surd && 0.08 \pm
0.05\% &&\omit\cr\space
\tablerule\space
& $c^{8,1}_{35}$ && (N_1 +2\Sigma_1 +\Xi_1) +2(\Delta_1
+2\Sigma^*_1 +\Xi^*_1) && 1 && \epsilon^\prime && && 0.08 \pm 0.13\%
&&\omit\cr\space
& $c^{8,1}_{405}$ && (-N_1 +10\Sigma_1 +11\Xi_1
+8\sqrt{3} \,\Lambda\Sigma^0) - 2(\Delta_1 +2\Sigma^*_1 +\Xi^*_1) && 1
&& \epsilon^\prime && && \mbox{---} &&\omit\cr\space
& $c^{27,1}_{405}$ && (N_1 -\Xi_1 +2\sqrt{3}
\,\Lambda \Sigma^0) + 2 (\Delta_1
-3\Sigma^*_1 -4\Xi^*_1) && 1/\N && \epsilon^\prime\epsilon &&  &&
\mbox{---} &&\omit\cr\space
\tablerule\space
&  && 25(\Sigma_1 +\Xi_1) -3(4\Sigma^*_1 -3\Xi^*_1) && 1 &&
\epsilon^\prime && && 0.36 \pm 0.02\% &&\omit\cr\space
&  && N_1 - \Xi_1 && 1 && \epsilon^\prime && && 0.23 \pm 0.03\% &&
\omit\cr\space
&  && 5(2N_1 + \Sigma_1 -\Xi_1) -3(4\Sigma^*_1 -3\Xi^*_1) && * && *
&& && 0.005 \pm 0.018\% &&\omit\cr\space
&  && \Sigma^*_1 -2\Xi^*_1 && 1/\N && \epsilon^\prime\epsilon &&
&& 0.04 \pm 0.03\% &&\omit\cr\space
\tablerule\space\tablerule\space
&    && \omit \hfil $I=2$ \hfil &&   &&  &&
&&\omit&&\omit\cr
\tablerule\space\tablerule\space
& $c^{27,2}_{(2)}$ && 35\Sigma_2 -2(3\Delta_2
+\Sigma^*_2) && 1/\N && \epsilon^{\prime\prime} && No && 0.12 \pm
0.03\% &&\omit\cr\space
& $c^{27,2}_{(3)}$ && 7\Sigma_2
-(3\Delta_2 +\Sigma_2^*) && 1/\N && \epsilon^{\prime\prime} &&\surd
&& 0.16 \pm 0.06\% &&\omit\cr\space
& $c^{64,2}_{(3)}$ && \Delta_2 -2 \Sigma_2^* &&
1/\N^2 && \epsilon^{\prime\prime}\epsilon &&\surd && 0.20 \pm 0.09\%
&&\omit\cr\space
\tablerule\space
& $c^{27,2}_{405}$ && \Sigma_2 +2(3\Delta_2
+\Sigma_2^*) && 1/\N && \epsilon^{\prime\prime} && && 0.26 \pm 0.15\%
&&\omit\cr\space
\tablerule\space\tablerule\space
&    && \omit \hfil $I=3$ \hfil &&   &&  &&
&&\omit&&\omit\cr
\tablerule\space\tablerule\space
& $c^{64,3}_{(3)}$ && \Delta_3 && 1/\N^2 &&
\epsilon^{\prime\prime}\epsilon^\prime &&\surd && 0\% (fixed)
&&\omit\cr\space
\space\tablerule\space\tablerule
}}}
\end{table}

\vfill\break\eject

\narrowtext


\begin{figure}
\caption{$SU(2F)$ spin-flavor representation for ground-state baryons.
The Young tableau has $\N$ boxes.}
\label{fig:groundstate}
\end{figure}

\begin{figure}
\caption{$SU(F)$ flavor representations for the tower of baryon states
with $J=\frac 1 2$, $\frac 3 2$, $\ldots$, $\frac \N 2$.  Each Young
tableau has $\N$ boxes.}
\label{fig:flavorreps}
\end{figure}

\begin{figure}
\caption{Weight diagram for the $SU(3)$ flavor representation of
the spin-${1 \over 2}$ baryons. The long side of the weight diagram
contains ${1 \over 2}\left( \N + 1 \right)$ weights. The numbers
denote the multiplicity of the weights.}
\label{fig:weight1/2}
\end{figure}

\begin{figure}
\caption{Weight diagram for the $SU(3)$ flavor representation of
the spin-${3 \over 2}$ baryons. The long side of the weight diagram
contains ${1 \over 2}\left( \N - 1 \right)$ weights. The numbers
denote the multiplicity of the weights.}
\label{fig:weight3/2}
\end{figure}

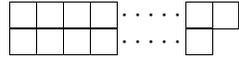
\vfill\break\eject

\def\sqr{\mathchoice\ssqr8{.4}\ssqr8{.4}\ssqr{5}{.3}\ssqr{4}{.3}}
\def\bsqr{\ssqr{10}{.1}}
\def\nbox{\hbox{$\bsqr\bsqr\bsqr\bsqr\raise2.7pt\hbox{$\,\cdot\cdot
\cdot\cdot\cdot\,$}\bsqr\bsqr\bsqr$}}

\centerline{$$\nbox$$}
\bigskip
\centerline{Figure 1}
\vskip1in

\def\nboxA{\vbox{\hbox{$\bsqr\bsqr\bsqr\bsqr\raise2.7pt\hbox{$\,\cdot
\cdot\cdot\cdot\cdot\,$}\bsqr\bsqr$}\nointerlineskip
\kern-.3pt\hbox{$\bsqr$}}}

\def\nboxE{\vbox{\hbox{$\bsqr\bsqr\bsqr\raise2.7pt\hbox{$\,\cdot\cdot
\cdot\cdot\cdot\,$}\bsqr\bsqr\bsqr\bsqr$}\nointerlineskip
\kern-.2pt\hbox{$\bsqr\bsqr\bsqr\raise2.7pt\hbox{$\,\cdot\cdot\cdot
\cdot\cdot\,$}\bsqr$}}}

\def\nboxF{\vbox{\hbox{$\bsqr\bsqr\bsqr\bsqr\raise2.7pt\hbox{$\,\cdot
\cdot\cdot\cdot\cdot\,$}\bsqr\bsqr$}\nointerlineskip
\kern-.2pt\hbox{$\bsqr\bsqr\bsqr\bsqr\raise2.7pt\hbox{$\,\cdot\cdot
\cdot\cdot\cdot\,$}\bsqr$}}}

\centerline{$$\nboxF$$}
\smallskip
\centerline{$J= \frac 1 2$}
\bigskip
\centerline{$$\nboxE$$}
\smallskip
\centerline{$J= \frac 3 2$}
\bigskip
\centerline{$\cdot$}
\centerline{$\cdot$}
\centerline{$\cdot$}
\bigskip
\centerline{$$\nbox$$}
\smallskip
\centerline{$J= \frac \N 2$}
\bigskip
\centerline{Figure 2}
\vskip1in

\def\onedot{\makebox(0,0){$\scriptstyle 1$}}
\def\twodot{\makebox(0,0){$\scriptstyle 2$}}
\def\threedot{\makebox(0,0){$\scriptstyle 3$}}
\def\fourdot{\makebox(0,0){$\scriptstyle 4$}}

\setlength{\unitlength}{3mm}

\centerline{\hbox{
\begin{picture}(20.79,18)(-10.395,-8)
\multiput(-1.155,10)(2.31,0){2}{\onedot}
\multiput(-2.31,8)(4.62,0){2}{\onedot}
\multiput(-3.465,6)(6.93,0){2}{\onedot}
\multiput(-4.62,4)(9.24,0){2}{\onedot}
\multiput(-5.775,2)(11.55,0){2}{\onedot}
\multiput(-6.93,0)(13.86,0){2}{\onedot}
\multiput(-8.085,-2)(16.17,0){2}{\onedot}
\multiput(-9.24,-4)(18.48,0){2}{\onedot}
\multiput(-10.395,-6)(20.79,0){2}{\onedot}
\multiput(-9.24,-8)(2.31,0){9}{\onedot}
\multiput(0,8)(2.31,0){1}{\twodot}
\multiput(-1.155,6)(2.31,0){2}{\twodot}
\multiput(-2.31,4)(2.31,0){3}{\twodot}
\multiput(-3.465,2)(2.31,0){4}{\twodot}
\multiput(-4.62,0)(2.31,0){5}{\twodot}
\multiput(-5.775,-2)(2.31,0){6}{\twodot}
\multiput(-6.93,-4)(2.31,0){7}{\twodot}
\multiput(-8.085,-6)(2.31,0){8}{\twodot}
\end{picture}
}}
\bigskip
\centerline{Figure 3}
\vskip1in

\centerline{\hbox{
\begin{picture}(20.79,18)(-8.085,-8)
\multiput(-1.155,10)(2.31,0){4}{\onedot}
\multiput(-2.31,8)(9.24,0){2}{\onedot}
\multiput(-3.465,6)(11.55,0){2}{\onedot}
\multiput(-4.62,4)(13.86,0){2}{\onedot}
\multiput(-5.775,2)(16.17,0){2}{\onedot}
\multiput(-6.93,0)(18.48,0){2}{\onedot}
\multiput(-8.085,-2)(20.79,0){2}{\onedot}
\multiput(-6.93,-4)(18.48,0){2}{\onedot}
\multiput(-5.775,-6)(16.17,0){2}{\onedot}
\multiput(-4.62,-8)(2.31,0){7}{\onedot}
\multiput(0,8)(2.31,0){3}{\twodot}
\multiput(-1.155,6)(6.93,0){2}{\twodot}
\multiput(-2.31,4)(9.24,0){2}{\twodot}
\multiput(-3.465,2)(11.55,0){2}{\twodot}
\multiput(-4.62,0)(13.86,0){2}{\twodot}
\multiput(-5.775,-2)(16.17,0){2}{\twodot}
\multiput(-4.62,-4)(13.86,0){2}{\twodot}
\multiput(-3.465,-6)(2.31,0){6}{\twodot}
\multiput(1.155,6)(2.31,0){2}{\threedot}
\multiput(0,4)(4.62,0){2}{\threedot}
\multiput(-1.155,2)(6.93,0){2}{\threedot}
\multiput(-2.31,0)(9.24,0){2}{\threedot}
\multiput(-3.465,-2)(11.55,0){2}{\threedot}
\multiput(-2.31,-4)(2.31,0){5}{\threedot}
\multiput(2.31,4)(2.31,0){1}{\fourdot}
\multiput(1.155,2)(2.31,0){2}{\fourdot}
\multiput(0,0)(2.31,0){3}{\fourdot}
\multiput(-1.155,-2)(2.31,0){4}{\fourdot}
\end{picture}
}}
\bigskip
\centerline{Figure 4}

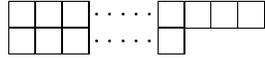
\end{document}